\documentclass[10pt,journal,compsoc]{IEEEtran}

\ifCLASSOPTIONcompsoc
  \usepackage[nocompress]{cite}
\else
  \usepackage{cite}
\fi

\ifCLASSINFOpdf
\else
\fi

\usepackage[justification=justified]{caption}
\usepackage[tight,footnotesize]{subfigure}
\usepackage{color}
\usepackage{graphicx}
\usepackage{amsmath}
\usepackage{amsmath}
\usepackage{amsopn}
\usepackage{subfig}
\usepackage{subfloat}
\usepackage{graphicx}
\usepackage{subfigure} 
\usepackage{algorithm, algorithmic}

\DeclareMathOperator*{\argmin}{arg\,min}

\begin{document}
\title{DMRO: A Deep Meta Reinforcement Learning-based Task Offloading Framework \\for Edge-Cloud Computing}

\author{Guanjin~Qu
        and~Huaming~Wu,~\IEEEmembership{Member,~IEEE}% <-this % stops a space
\IEEEcompsocitemizethanks{\IEEEcompsocthanksitem G. Qu and H. Wu are with the Center for Applied Mathematics, Tianjin University, Tianjin 300072, China.\protect\\
E-mail: \{guanjinqu, whming\}@tju.edu.cn
}% <-this % stops an unwanted space
\thanks{(Corresponding author: Huaming Wu)}}


\markboth{}%
	{Qu \MakeLowercase{\textit{et al.}}: DMRO: A Deep Meta Reinforcement Learning-based Task Offloading Framework for Edge-Cloud Computing}




\IEEEtitleabstractindextext{%
\begin{abstract}
With the continuous growth of mobile data and the unprecedented demand for computing power, resource-constrained edge devices cannot effectively meet the requirements of Internet of Things (IoT) applications and Deep Neural Network (DNN) computing. As a distributed computing paradigm, edge offloading that migrates complex tasks from IoT devices to edge-cloud servers can break through the resource limitation of IoT devices, reduce the computing burden and improve the efficiency of task processing. However, the problem of optimal offloading decision-making is NP-hard, traditional optimization methods are difficult to achieve results efficiently. Besides, there are still some shortcomings in existing deep learning methods, e.g., the slow learning speed and the failure of the original network parameters when the environment changes. To tackle these challenges, we propose a Deep Meta Reinforcement Learning-based Offloading (DMRO) algorithm, which combines multiple parallel DNNs with Q-learning to make fine-grained offloading decisions. By aggregating the perceptive ability of deep learning, the decision-making ability of reinforcement learning, and the rapid environment learning ability of meta-learning, it is possible to quickly and flexibly obtain the optimal offloading strategy from the IoT environment. Simulation results demonstrate that the proposed algorithm achieves obvious improvement over the Deep Q-Learning algorithm and has strong portability in making real-time offloading decisions even in time-varying IoT environments.
\end{abstract}

}

\maketitle

\IEEEdisplaynontitleabstractindextext

% \fi
\IEEEpeerreviewmaketitle

\IEEEraisesectionheading{\section{Introduction}\label{sec:introduction}}

\IEEEPARstart {W}{ith} the development of Internet of Things (IoT) and communication technologies, a large number of computation-intensive tasks need to be transferred from IoT devices to the cloud server for execution \cite{ai2018edge}. However, the task offloading process usually involves large amounts of data transmission, which will result in high latency for IoT applications. The emergence of Mobile Edge Computing (MEC) can effectively alleviate this challenge. As a distributed computing paradigm, edge offloading that migrates complex tasks from IoT devices to edge-cloud servers can provide computing services for edge caching, edge training, and edge inference \cite{Xu2020ASO}. Before the IoT application being offloaded to the cloud server, it needs to pass through the edge server, such as the base station. The edge server is closer to the device than the cloud server, so it has greater bandwidth and response time. By utilizing the computing and decision-making capabilities of the edge server, the task computing of the device can be offloaded to different servers, thereby reducing computing latency and energy consumption \cite{Hameed2014,wu2018energy}.

The process of  task offloading will be affected by different factors, such as user habits, wireless channel communication, connection quality, mobile device availability and cloud server performance. Therefore, making the optimal decision is the most critical issue for edge offloading. It needs to decide whether the task should be offloaded to the edge server or cloud server. If a large number of tasks are offloaded to the cloud server, the bandwidth will be occupied, which will greatly increase the transmission delay. Therefore, we need to have a reasonable offloading decision scheme so that it can reasonably allocate each task to the processing server. On the one hand, there are a large number of repetitive or similar tasks in the IoT environment, which often need to be retrained from scratch, resulting in inefficient offloading decision-making; on the other hand, some IoT application scenarios have strict time constraints on task decision-making, and the slow learning speed of Convolutional Neural Network (CNN) is not suitable to meet the requirements of resource heterogeneity and real-time in the MEC system.

Facing with the rapidly changing IoT application scenarios, we cannot readjust the task offloading decision and wireless resource allocation through recalculation every time the MEC environment changes, otherwise, it will cause higher service delay and cost \cite{mao2017survey}. Although some good results have been achieved in offloading decision-making of MEC by introducing intelligent algorithms such as deep reinforcement learning, there are still challenges such as slow learning speed, and failure of original network parameters when the model environment changes. In practical application scenarios, the MEC environment is often affected by many factors anytime and anywhere. Conventional intelligent algorithms are usually based on neural networks. When the MEC environment changes, its original parameters will all fail and a large amount of training data is required to train from scratch, which makes the learning efficiency low. Such repeated training will consume resources and weaken the performance of the MEC system. At the same time, in order to improve efficiency, high configuration equipment is also required to adapt to high-intensity training.

Considering the delay and energy consumption of IoT, offloading decisions can be made for a workflow with a series of dependent tasks. However, this problem is NP-hard, traditional optimization methods are difficult to achieve results efficiently. 
One promising way of addressing the above issue is to bring deep learning techniques, such as Deep Reinforcement Learning (DRL), into the computing paradigm of edge-cloud collaboration. Unfortunately, conventional DRL algorithms have the disadvantage of slower learning speed, which is mainly due to the weak inductive bias. A learning procedure with weak inductive bias will be able to adapt to a wide range of situations, however, it is generally less efficient~\cite{botvinick2019reinforcement}.

In this paper, we design an edge-cloud offloading framework with a cloud server, one edge server, and multiple IoT devices, where IoT devices can choose to shift their computing tasks either to edge servers or cloud servers. Edge servers make offloading decisions based on task information for each device, reducing latency and energy consumption. We propose an efficient offloading decision-making method based on deep meta reinforcement learning \cite{wang2016learning} that takes advantage of DRL and meta-learning. To solve the problem of poor neural network portability, we introduce meta-learning to ensure that the offloading decision model can quickly adapt to the new environment by learning the initial parameters of the neural network. The main contributions of this study are summarized as follows:

\begin{itemize}
\item Formalizing the task placement problem for dependent tasks in MEC as a multi-objective optimization problem. To jointly minimize the delay and energy consumption of IoT, we propose an effective and efficient offloading framework with intelligent decision-making capabilities.

\item Proposing a Deep Meta Reinforcement learning-based Offloading (DMRO) framework that combines multiple parallel Deep Neural Networks (DNNs) and deep Q-learning algorithms to make offloading decisions. By aggregating the perceptive ability of deep learning, the decision-making ability of reinforcement learning, and the rapid environment learning ability of meta-learning, it is possible to quickly and flexibly obtain the optimal offloading strategy from the dynamic environment.

\item Aiming at the change of MEC environments, an initial parameter training algorithm based on meta-learning is proposed, where meta-learning is applied to solve the problem of poor portability of neural networks. By learning the initial parameters of the neural network under various network environments, the offloading decision model can quickly adapt to the new environment.
\end{itemize}

The rest of the paper is organized as follows. In Section
\ref{sec:relatedwork}, we review the related work. The system model and problem formulation are presented in Section \ref{sec:sysmodel}. The proposed Deep Meta Reinforcement learning-based Offloading (DMRO) framework is demonstrated in Section \ref{sec:edgeoffloading}. Section \ref{sec:evaluation} contains the
simulation and its results. Finally, Section \ref{sec:conclusion} concludes the paper and draws future works.

\section{Related work}
\label{sec:relatedwork}
MEC is an emerging computing paradigm, which can connect IoT devices to cloud computing centers through edge servers close to the device, thereby forming this task offloading mode in the IoT-edge-cloud computing environment. Among them, the cloud center is responsible for providing flexible and on-demand computing resources for the execution of mobile applications, and the edge server is responsible for deciding which computing tasks need to be offloaded and providing a limited amount of computing resources. Thus, the energy consumption of the device and computing delay of the application can be reduced. In general, the task offloading process includes the following key components:
\begin{itemize}
\item \emph{Application Partition}: Since different tasks usually have different amounts of computation and communication, before performing task offloading operation, it is better to divide the task into a workflow with multiple associated subtasks or as a series of independent subtasks \cite{Zhang2019Automatic}, and then offload the subtasks separately. Among them, some subtasks are executed on the IoT devices, the others are executed on the relatively powerful server, making full use of the server resources, thereby greatly reducing the load of the IoT devices and improving its endurance \cite{Zhou2020,wu2019efficient}.

%An application can be divided into multiple tasks, then we need to divide before the task is offloaded, we can divide a task into a workflow with multiple associated subtasks, or as a series of independent subtasks, \cite{wu2019efficient} Provides a method of task division.

\item \emph{Resource Allocation}: After the offloading decision is made, resources need to be allocated, including computing power, communication bandwidth, and energy consumption.
\end{itemize}

At present, task offloading algorithms related to decision-making can be divided into traditional methods and intelligent algorithms using artificial intelligence \cite{mao2017survey}.

\subsection{Traditional Offloading Decision-Making}
Due to the NP-hardness of offloading decision problems in MEC, when the number of tasks increases, it is easy to encounter problems such as computational explosion. A diversity of platforms and algorithms \cite{Liu2020SCHED,Liang2020Joint,Liu2020MATCH,Xie2019novel,Peng2020End} have been proposed to solve the optimization problems of offloading binary decisions in edge-cloud environments.
%A traditional way is to use Lyapunov optimization. 

A Lyapunov optimization framework was proposed in \cite{li2019lyapunov} to weigh the offloading system and the queue backlog. eTime \cite{shu2013etime} was a cloud-to-device energy-efficient data transmission strategy based on Lyapunov-optimization, with more focus on data transmission optimization. Other references using Lyapunov optimization for offloading decision-making can be found in \cite{mao2016dynamic,zhang2018energy-delay,wu2018energy}. Markov processes and 
queueing models have been also widely applied for making offloading decision. The offloading approach proposed in \cite{Wu2017Stochastic} supported two delayed offloading policies, i.e., a partial offloading model where jobs can leave the slow offloading to be executed locally, and a full offloading model where jobs can be offloaded directly via the cellular network. Besides, a computing offloading game theory has been developed in \cite{li2018computing}, which proposed C-SGA (a fast Stackelberg game algorithm) and F-SGA (a complex Stackelberg game algorithm) to solve the decision problem of IoT-enabled cloud-edge computing. However, these optimization-based offloading algorithms can only obtain results after multiple iterations, which often involve too many complex calculation operations.

Conventional task offloading techniques usually apply some heuristic algorithms. A particle swarm optimization-based offloading decision algorithm was given in \cite{yiqing2007improved}. A computing method called COM was proposed in \cite{xu2019computation} to solve the problem of computation offloading in the cloud environment. Goudarzi et al.\cite{goudarzi2020application} gave a genetic algorithm that can solve the task offload problem in a multi-user multi-cloud multi-edge environment. However, heuristic algorithms are still difficult to solve complex problems that require a large amount of computation, and additional computation is also introduced, which results in high running time cost in offloading decision-making.  

%\cite{wu2019efficient}

% Many optimization-based algorithms, e.g., traversal or linear programming, can only obtain results after multiple iterations, which often involve too many complex calculation operations, e.g., matrix inversion and singular value decomposition, resulting in high running time cost in offloading decision-making.
 
\subsection{Intelligent Offloading Decision-Making}

With the rapid development of computer science and the popularization of Artificial Intelligence (AI), deep learning has begun to be applied to solve the problem of offloading decision-making. Edge intelligence \cite{Xu2020ASO} or intelligent edge \cite{Yu2020Intelligent}, that is, the convergence of edge computing and AI, takes advantage of both to achieve mutual benefit \cite{wang2020convergence}. On the one side, optimizing DNNs through task offloading has become a new direction in edge intelligence research since edge computing can offload complex computing tasks to edge/cloud servers. On the other side, deep learning-driven approaches can facilitate offloading decision making, dynamic resource allocation and content caching, benefit in coping with the growth in volumes of communication and computation for emerging IoT applications \cite{cao2019intelligent}.

Classic AI methods including deep learning and reinforcement learning, can provide more reasonable and intelligent solutions to solve the offloading decision problem in edge computing. Deep learning methods refer to the classification of the input task information through the multi-layer neural network to determine the final offloading position. Huang et al.~\cite{huang2018distributed} provided an algorithm that adopted distributed deep learning to solve the offloading problem of mobile edge networks. It used parallel and distributed DNNs \cite{Castello2019} to produce offloading decisions and achieved good results. A hybrid offloading model with the collaboration of Mobile Cloud Computing (MCC) and MEC was established in \cite{Wu2020Collaborate}, where a distributed deep learning-driven task offloading (DDTO) algorithm was proposed to generate near-optimal offloading decisions over the IoT devices, edge cloud server, and central cloud server. Besides, Neurosurgeon~\cite{kang2017neurosurgeon} was a fine-grained partitioning method that can find the optimal dividing point in DNNs according to different factors, and made full use of the resources of cloud servers and mobile devices to minimize the computational delays or energy consumption in IoT environments.

 In some cases, however, it is still difficult to treat task offloading decision-making as a classification problem to be solved by using deep learning techniques, which are mostly supervised learning. In addition, it is difficult to find labeled training sets for training on offloading decision problems. Reinforcement learning, as one of the paradigms of machine learning, is used to solve the interaction between the agent and the environment through learning, so as to achieve maximum return or specific goals. An edge-cloud task offloading framework using a Deep Imitation Learning (DIL) \cite{Zhang2018Deep} was proposed in \cite{Yu2020Intelligent}, while training DNN model with DIL is still computation-intensive. Deep Reinforcement Learning (DRL) methods that combined with neural network and reinforcement learning can be used to solve the task offloading decision problem in the MEC environment. The final decision is the maximum reward action under the interaction with the environment. The premise of using DRL algorithms for task offloading decision is that it can be regarded as a Markov process, in which three spaces named state, action, and reward are 
established.
 Among them, the task information is input into the state, and the offloading decision is located in the action space. Zhang et al.~\cite{zhang2018joint} proposed an offloading decision scheme based on Actor-Critic algorithm. In \cite{huang2018deep} and \cite{huang2019deep}, task offloading decisions were made based on DRL algorithms, e.g., Deep Q-Learning Network (DQN) and Double Deep Q-learning Network (DDQN)-based algorithms, however, the cloud server was not considered in the MEC environment and they usually require to learn from scratch when the environment changes. 

Currently, the role of DRL is to choose an optimal edge computing environment or location for the current task according to its status and environment. However, each time the IoT environment changes, the offloading decision has to be recalculated, which leads to more service delays and higher costs. In addition, DRL algorithms are still limited with slower learning speed and are generally less efficient in solving the offloading decision-making problem~\cite{botvinick2019reinforcement}.
Therefore, it is necessary to find an intelligent method that can learn knowledge and quickly provide better offloading decisions with the change of environment. Unlike traditional machine learning that only trains a general learning model for edge offloading, the goal of meta-learning is learning to learn fast, that is, to make the model become a learner \cite{andrychowicz2016learning,gupta2018meta}. After completing multiple learning tasks, it can quickly complete new learning tasks by learning prior knowledge or exploring learning strategies \cite{finn2017model}. Therefore, it quickly adapts to complex and changing environments and can be used to improve the robustness of task offloading decisions in IoT environments.

\section{System Model and Problem Formulation}
\label{sec:sysmodel}
In this section, we give an overview of the system model and then define the delay model and energy consumption model. On this basis, the optimization problem of computation offloading is formulated.

\subsection{System Model}
\begin{figure}[htbp]
\centering
\includegraphics[width=3.6in, trim=0.2in 0.4in 0.2in 0.25in]{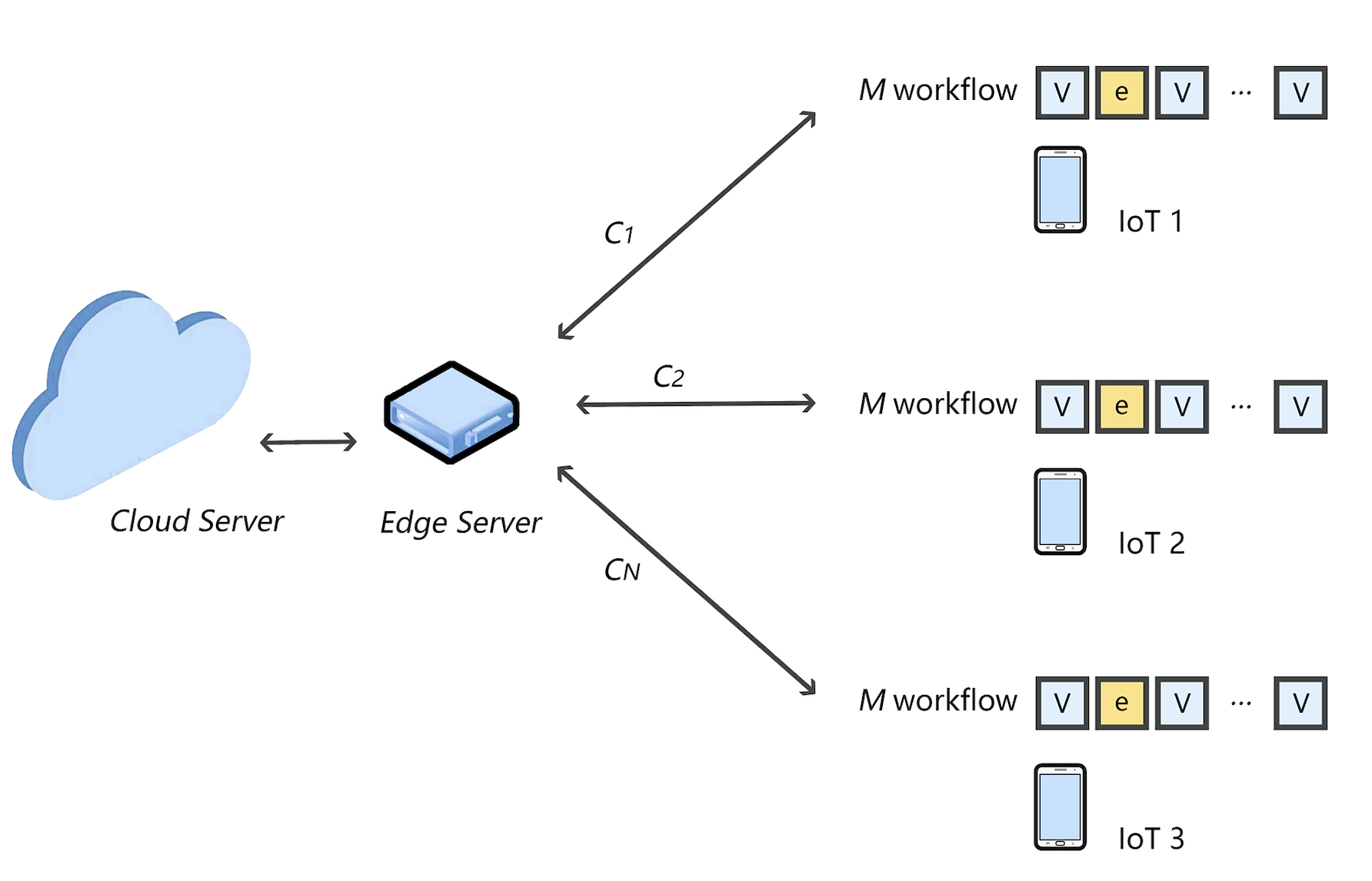}
\caption{System model of edge-cloud computing with multiple IoT devices}
\label{system_model}   
\end{figure}

The system model for task offloading in IoT-edge-cloud computing environments is shown in Fig.~\ref{system_model}. The proposed framework is composed of a cloud server, an edge server, and multiple IoT devices, where the IoT devices can either execute locally or offload their workflow to the cloud server or edge server. 

In this framework, edge servers are distributed near the devices and have high bandwidth. The edge server accepts workflow information from the device and makes fine-grained offloading decisions. The program for each device can be divided into sequential workflows. We assume the $x$-th workflow is defined as follows:
\begin{equation}
R_{x}=
\left\{ 
v_{1},e_{1,2},v_{2},\cdots,v_{i},e_{i,j},v_{j},\cdots,e_{n-1,n},v_{n}
\right\},
\end{equation}
where $v_{i}$ denotes $i$-th task in the workflow, and $e_{i,j}$ illustrates the set of data flows between tasks $v_{i}$ and $v_j$.

Each workflow $x$ can determine whether to offload its task $v_{i}$ or not, and the offloading decision is denoted by a Matrix variable:
\begin{equation}
b_{x, i} \in\left(b_{0}, \quad b_{1}, \quad b_{2}\right),
\end{equation}
specifically, $b_{0}=[1 \quad  0 \quad 0]^T$, $b_{1}=[0 \quad  1 \quad 0]^T$ and $b_{2}=[0 \quad  0 \quad 1]^T$ denote the decision that workflow $x$ to execute its $i$-th task locally, offload $i$-th task to the edge server, and offload $i$-th task to the cloud server, respectively.

%$b_{0}=\left[\begin{array}{l}1 \\ 0 \\ 0\end{array}\right]$
%$b_{0}=\left[\begin{array}{l}0 \\ 1 \\ 0\end{array}\right]$
%$b_{0}=\left[\begin{array}{l}0 \\ 0 \\ 1\end{array}\right]$

\subsection{Delay Model}

The delay caused by computation offloading includes computation delay and transmission delay. We do not consider the delay incurred in offloading decision-making because the time required to make the decision is short. Therefore, the computational delay of task $v_i$ is calculated by:
\begin{equation}
T_{i}^{c}=\left\{\begin{array}{ll}\frac{v_{i}}{C_{0}}, & b_{x, i}=b_{0}, \\ \frac{v_{i}}{C_{1}}, &  b_{x, i}=b_{1}, \\ \frac{v_{i}}{C_{2}}, &  b_{x, i}=b_{2},\end{array}\right.
\end{equation}
where $C_0$, $C_1$ and $C_3$ stand for the computing power of the IoT, the computing power of the edge server and the computing power of the cloud server, respectively.

The transmission delay between tasks $v_i$ and $v_j$ is:
\begin{equation}
T_{i, j}^{t}=\left\{\begin{array}{ll}0, & b_{x, i}=b_{x, j}, \\ \frac{e_{i, j}}{B_{0,1}}, & b_{x, i}=b_{0}, b_{x, j}=b_{1} \text { or } b_{x, i}=b_{1}, b_{x, j}=b_{0}, \\ \frac{e_{i, j}}{B_{1,2}}, & b_{x, i}=b_{1}, b_{x, j}=b_{2} \text { or } b_{x, i}=b_{2}, b_{x, j}=b_{1}, \\ \frac{e_{i, j}}{B_{0,2}}, & b_{x, i}=b_{0}, b_{x, j}=b_{2} \text { or } b_{x, i}=b_{2}, b_{x, j}=b_{0},\end{array}\right.
\end{equation}
where $B_{0,1}$ denotes the allocated bandwidth between the IoT device and the edge server. $B_{1,2}$ is the allocated bandwidth between the cloud server and the edge server. Similarly, we denote $B_{0,2}$ as the allocated bandwidth between the IoT device and the cloud server. 

The total delay for workflow $x$ is calculated as:
\begin{equation}
T_{x}=\sum_{i=1}^{N}\left(T_{i}^{c}+T_{i, i+1}^{t}\right),
\end{equation}
where the workflow $x$ has $N$ associated tasks.

\subsection{Energy Consumption Model}
The energy consumption model of workflow $x$ can be expressed as:
\begin{equation}
E_{x}=E_{x}^{\text {local}}+\alpha E_{x}^{\text {edge}}+\beta E_{x}^{\text {cloud}},
\end{equation}
where $\alpha$ and $\beta$ are weights of the energy consumption at the edge server and at the cloud server, respectively. When $\alpha$ = $\beta = 0$, we only consider the energy consumption at the IoT device. For simplicity, we ignore the energy consumed during task transmission.

The energy consumption of task $v$ is calculated as:
\begin{equation}
E_{i}=\left\{\begin{array}{ll}v_{i}\cdot d_{local}, &  b_{x, i}=b_{0}, \\ v_{i} \cdot d_{edge}, &  b_{x, i}=b_{1}, \\ v_{i} \cdot d_{cloud}, &  b_{x, i}=b_{2},\end{array}\right.
\end{equation}
where $d_{local}$, $d_{edge}$ and $d_{cloud}$ denote the local energy consumption per data bit, the edge energy consumption per data bit and the cloud energy consumption per data bit, respectively.

Therefore, the energy consumption model of workflow $x$ can be expressed by:
\begin{equation}
E_{x}=\sum_{i=1}^{N}\left[E_{i}, \alpha E_{i}, \beta E_{i}\right] \cdot b_{x, i}.
\end{equation}

\subsection{Problem Formulation}
To minimize both the delay for completing all workflows and the corresponding energy consumption simultaneously, we first introduce a system utility $Q(x, b)$, which is defined as the weighted sum of energy consumption and workflow completion delay, as follows:
\begin{align}
Q(x, b)&=\sum_{x=1}^{M}\left(T_{x}+\delta E_{x}\right)\nonumber\\
&=\sum_{x=1}^{M}\left(\sum_{i=1}^{N}\left(T_{i}^{c}+T_{i, i+1}^{t}\right)+\delta \sum_{i=1}^{N}\left[E_{i}, \alpha E_{i}, \beta E_{i}\right] b_{x, i}\right),
\end{align}
where there are $M$ workflows in total, each workflow has $N$ associated tasks, and $\delta$ denotes the weight of energy consumption and task completion time. 

The optimization problem can be formulated as follows:

\begin{align}
&\min _{b}  Q(x, b),\\
s.t.: \;&b_{x, i} \in\left(\left[\begin{array}{l}1 \\ 0 \\ 0\end{array}\right],\left[\begin{array}{l}0 \\ 1 \\ 0\end{array}\right],\left[\begin{array}{l}0 \\ 0 \\ 1\end{array}\right]\right).
\end{align}

\section{Deep Meta Reinforcement Learning-based Offloading Framework}
\label{sec:edgeoffloading}
\begin{figure}[ht!]
	\centering
	\includegraphics[width=3.7in, trim=0.8in 1.5in 0in 0.5in]{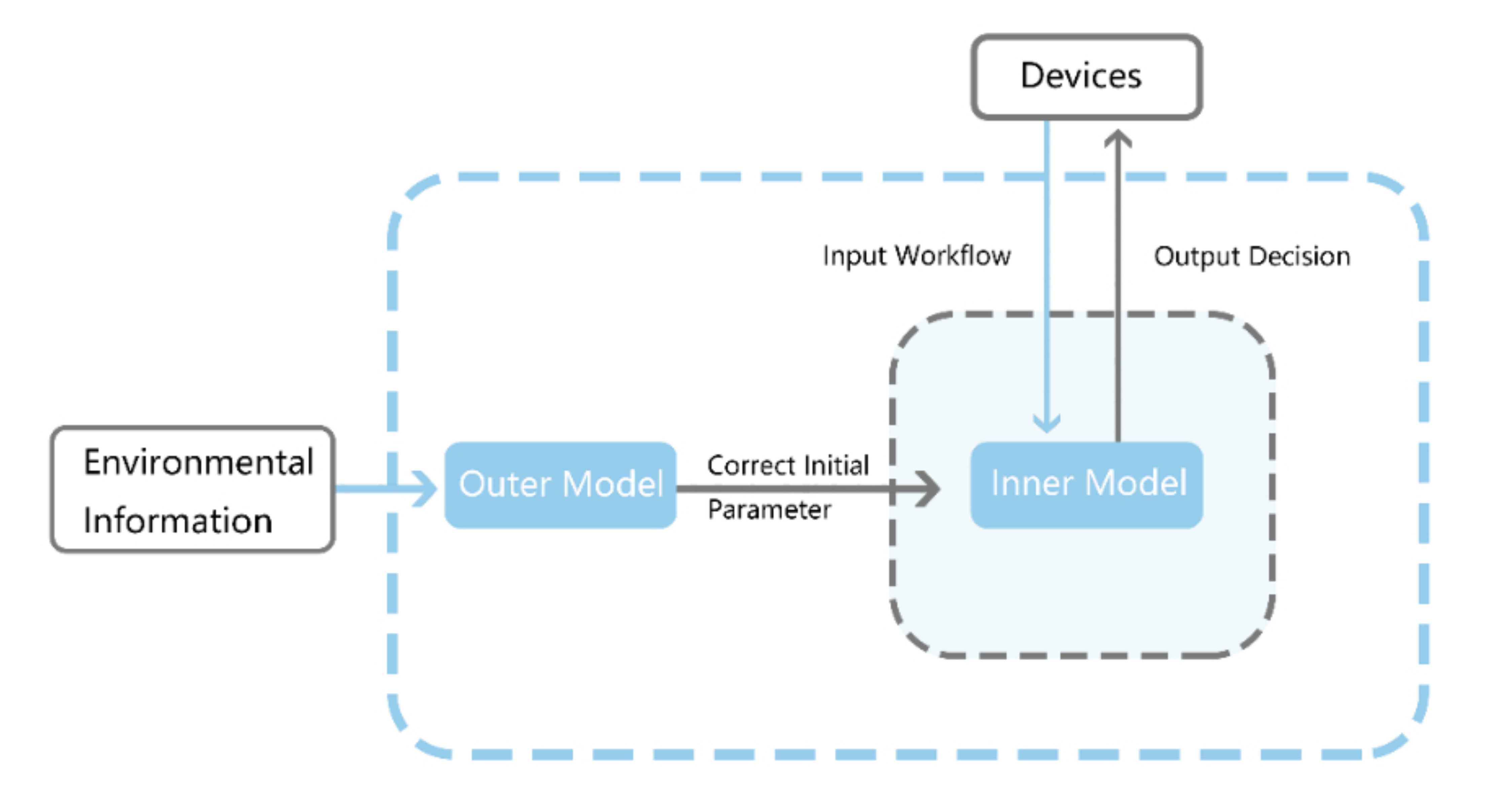}
	\caption{The proposed deep meta reinforcement learning-based offloading framework}
	\label{Architecture}   
\end{figure}

 To effectively solve the optimization problem defined in (10), we then propose a Deep Meta Reinforcement learning-based Offloading (DMRO) framework as shown in Fig.~\ref{Architecture}, where a series of dependent tasks are considered comprehensively, in order to give a specific offloading decision for each task. The proposed learning-driven offloading framework contains a task offloading decision model based on distributed reinforcement learning algorithm and a training model based on meta-learning, aiming to solve the problem of poor portability of neural networks.

The DMRO framework can be divided into two layers of models. The inner model is an offloading decision model based on a distributed deep reinforcement learning, which is responsible for receiving the workflow and training the model parameters to give the final offloading decisions for different tasks. The outer model is the meta-learning part, which is responsible for training the initial parameters to improve the portability of the model. When the environment of the MEC system changes, such as the performance of the edge server or the bandwidth between the IoT device and the edge server, it can adjust the parameters of the neural network in the inner model, so that the system can quickly adapt to the new environment. When the workflow is input into the edge offloading system, the outer model first determines whether the external environment has changed, in order to determine whether to adjust the initial parameters. After that, the workflow will enter the inner model, which will make the offloading decision, and store the state and action in memory for the training of the neural network.

 %The inner model is mainly reinforcement learning algorithms, which are responsible for giving specific offloading decisions for different tasks in the workflow. The outer model is the meta-learning algorithm, which is responsible for training the initial parameters to improve the portability of model. 

Furthermore, to increase the portability of the model, speed up the decision-making process and reduce the amount of computation, we design a deep meta-reinforcement learning-based method, which also combines the function of memory playback (replay memory), so that the decision-making system can adapt to the new environment quickly and give the optimal offloading decision when the environment changes. In addition, the generated offloading decisions are stored in the replay memory summary for further learning.

\begin{figure*}[ht!]
	\centering
	\includegraphics[width=6in, trim=0.4in 1.0in 0.4in 0.35in]{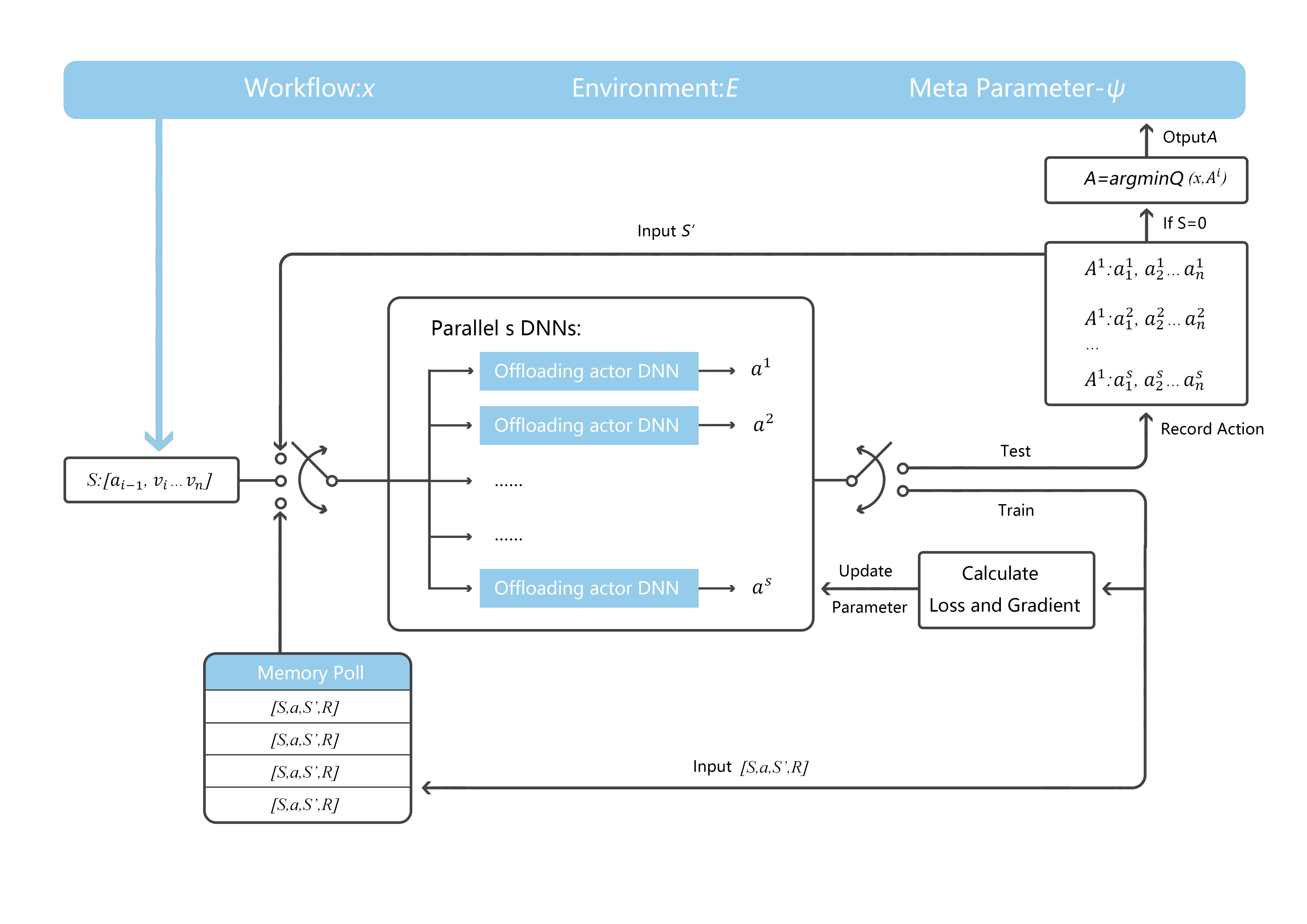}
	\caption{Illustration of the distributed deep reinforcement learning-based offloading scheme}
	\label{3}   
\end{figure*}  

\subsection{Inner Model}
As shown in Fig.~\ref{3}, the inner model is based on a parallel Deep Reinforcement Learning (DRL) algorithm. We apply a classic reinforcement learning method named Q-learning, in which we input environmental parameters, labeled initial parameters and workflow $x$ into the inner model. 
 
 We use $a_i$ to represent the offloading decision of the $i$-th subtask of the workflow,which is defined as:
 \begin{eqnarray}
a_i=\left\{ \begin{array}{ll}
0,& \textrm{if subtask $i$ is executed on IoT device}, \\
1,& \textrm{if subtask $i$ is offloaded to edge server},\\
2,& \textrm{if subtask $i$ is offloaded to cloud server},
\end{array}\right.
\label{x1}
\end{eqnarray}
where $a_i=0$, $1$, and $2$ indicate that the $i$-th subtask is executed locally on the IoT device, the edge server, and the cloud server, respectively. 

We represent $S_i$ as the state when processing the $i$-th subtask in the workflow:
\begin{equation}
S_{i}=\left[a_{i-1}, e_{i-1, i}, v_{i}, e_{i, i+1}, v_{i+1}, \cdots, e_{n-1, n}, v_{n}\right],\; i\geq 1,
\end{equation}
where $a_{i-1}$ represents the execution position of a subtask in the workflow, which is set as $0$ at the beginning. Then the state $S_i$ is input to the neural network to find the $Q$ value of each action in this state. 

Here we have $s$ distributed neural network units. Each neural network action unit is parallel, including two DNNs with the identical structure, one of which is the target network for parameter freezing. Parameter freezing means that the two networks have the same structure, but the parameters of the frozen network will not be iterated every time. When the other network learns a certain number of times, the parameters are copied to the frozen network. The purpose of using parameter freezing is to reduce the relevance of learning~\cite{mnih2013playing}.
Each neural network unit will give the selected action value according to its own $Q$ value calculated by the greedy algorithm. In addition, we define a local objective function:
\begin{align}
F\left(S_{i}, a\right) &=T_{i}^{c}+T_{i-1, i}^{t}+\delta E_{i}, \\
b_{x, i-1}&=a_{i-1},\\
b_{x, i}&=a_{i},
\end{align}
where $F(S_i,a)$ can be interpreted as the weighted sum of the delay and energy consumption for selecting action $a$ in state $S_i$. We compare $F(S_i,a)$ values generated by the actions selected by different DNNs as a measure of the effects of the actions selected by different DNNs. The action with the lowest $F$ is set as the optimal solution in the state $S_i$. 

For the reward function $R(S_i,a)$ in DRL, if the action is the action value of the optimal solution, the reward value is the negative value of the minimum optimization function; otherwise, the reward value is the negative value of the maximum function. Then we choose the action of the optimal solution as $a_i$, and update the state $S_i$ as:
\begin{equation}
S_{i+1}=\left[a_{i}, e_{i, i+1}, v_{i+1}, e_{i+1, i+2}, v_{i+2}, \cdots, e_{n-1, n}, v_{n}\right]
\end{equation}

The algorithmic process of the proposed parallel DRL algorithm is as demonstrated in \textbf{Algorithm}~\ref{alg:1}. 
\subsubsection{Training Phase}
In the training phase, we input $[S_i,a_i,R(S_i,a),S_{i+1}]$ calculated by the neural network into the memory, and then continue to input the updated workflow into the neural network for calculation until all subtasks of the workflow have been processed. 

After processing a certain number of workflows, e.g., five times, we will randomly extract $[S_i,a_i,R(S_i,a),S_{i+1}]$ from the memory for empirical playback. The purpose of this step is to eliminate the correlation generated by the associated states. Then we update the parameters of the network as follows:
\begin{equation}
Q\left(S_{i}, a_{i}\right) \leftarrow(1-\theta) Q\left(S_{i}, a_{i}\right)+\theta\Big(R\left(S_{i}, a\right)+\mu \max _{a^{\prime} \in A}Q\left(S_{i+1}, a^{\prime}\right)\Big)
\end{equation}
where $Q\left(S_{i}, a_{i}\right)$ represents the $Q$ value function, which is calculated by the neural network, $Q\left(S_{i}, a_{i}\right)$ represents the $Q$ part is calculated by the network with the latest parameters, and  $\max _{a^{\prime} \in A}Q\left(S_{i+1}, a^{\prime}\right)$ is calculated by the network with the frozen parameters.
The learning rate $\theta\in\left[ 0, 1\right] $ is the weight of the current offloading experience. The discount factor $\mu\in\left[ 0, 1\right]$ denotes the short view of the IoT device regarding the future reward.

\subsubsection{Decision-Making Phase}

In the decision-making phase, we will make fine-grained offloading decisions for IoT devices. First, we obtain the action value $a$ generated by each DNN and fill it into $s$ action sets $A$. Then, we input the updated state to the neural network, and continue to find the execution method of the next subtask until all the subtasks are assigned. At this time, $A^i$ represents the offloading scheme given by the $i$-th DNN network to the workflow $x$, and the scheme $A^i$ with the minimum $Q(x,b)$ value is the final scheme $A$ and output to the device.
 
\begin{algorithm}
	\renewcommand{\algorithmicrequire}{\textbf{Input:}}
	\renewcommand{\algorithmicensure}{\textbf{Output:}}
	\caption{Distributed deep reinforcement learning algorithm}
	\label{alg:1}
	\begin{algorithmic}[1]
		\REQUIRE Workflow $x$, Environment: $E$, Meta-parameter: $\psi$
		\ENSURE Optimal offloading decision $A$
		\STATE 	Initialize the $s$ DNNs with meta-parameter $\psi$
		\STATE  Empty the memory pool
		\FOR {$i=1,2,3,\cdots,N$}        
		\STATE Replicate state $S_i$ to all $s$ DNNs
		\STATE Generate $s$-th offloading action $a_i^j$  via $\epsilon$-greedy policy
		\FOR {$j=1,2,3,\cdots,s$}
		\STATE Input $a_i^j$ to decision set $A^j$ as: $a_i^1$, $a_i^2$, $\cdots$, $a_i^j$	        
		\STATE Evaluate the local objective function $F(S_i,a_j^j)$, generate reward $R(S_i,a_i^j)$
		\IF {train}
		\STATE  $a_i^1=a_i$
		\ELSE
		\STATE $a_i$ is $a_i^j$ in turn 
		\IF {$i==n$}
		\STATE 	Select $A$ according to  $\argmin_{A^j} Q(S_i,A^j)$
		\STATE Output $A$ as offloading decision
		\ENDIF
		\STATE Input $[S_i,a_i,R(S_i,a_i),S_{i+1}]$ to memory pool
		\ENDIF        
		\ENDFOR
		\IF {Add data to memory five times}
		\STATE Extract $[S_i,a_i,R(S_i,a_i),S_{i+1}]$ from memory at random
		\STATE Replicate state $S_i$ to all $s$ DNNs
		\STATE Evaluate the local objective function $F(S_i,a_i)$, generate reward $R(S_i,a_i)$
		\STATE Update the $s$ DNNs weights $\theta$
		\ENDIF
		\ENDFOR        
	\end{algorithmic}  
\end{algorithm}

\subsection{Outer Model}
In the outer model, we propose a meta algorithm to learn the initial parameters. 

Based on the original algorithm described in \cite{finn2017model}, i.e., an initial parameter algorithm for training different image classification networks, we propose a novel algorithm for learning initial parameters in order to adapt to the training method of reinforcement learning. We train our decision-making engine by leveraging the deep meta-learning method, and then make rapid offloading decisions through IoT-edge-cloud computing environments. The algorithmic process of the proposed meta algorithm is as listed in \textbf{Algorithm}~\ref{alg:2}.
%\begin{figure}[htbp]
%	\centering
%	\includegraphics[scale=0.2]{Fig4. Illustration of Meta algorithm for training initial parameters.jpg}
%	\caption{ Illustration of Meta algorithm for training initial parameters}
%	\label{4}   
%\end{figure}  

 The principle of our meta algorithm is to input the decision-making and execution results of the workflow in different environments into the training model. Each time the training model randomly selects training samples in one environment for learning, and randomly selects another environment after learning. The purpose of training sample learning is to ensure that the parameters trained by the model will not be too close to the optimal solution in a specific environment. We use the parameters trained in this way as the initial parameters of the inner model.
\begin{algorithm}
	\renewcommand{\algorithmicrequire}{\textbf{Input:}}
	\renewcommand{\algorithmicensure}{\textbf{Output:}}
	\caption{Meta algorithm}
	\label{alg:2}
	\begin{algorithmic}[1]
		\REQUIRE Workflow $x$, Environment: $E$
		\ENSURE Optimal offloading decision $A$
		\STATE 	Initialize the DNNs with parameter $\theta_0$
		\STATE  Empty the memory pool
		\FOR {$i=1,2,3,\cdots,N$}  
		\STATE Randomly select environment
		\STATE Input state $S_i$ to DNN
		\STATE Generate offloading action $a_i$ via $\epsilon$-greedy policy
		\STATE generate reward $R(S_i,a_i)$ via Random Environment
		\STATE Input $[S_i,a_i,R(S_i,a_i),S_{i+1}]$ to memory pool
		        \IF {Add data to memory five times}
		        \STATE Randomly select environment
		\STATE Extract $[S_i,a_i,R(S_i,a_i),S_{i+1}]$ from memory at random
		\STATE Replicate state $S_i$ to the DNNs
		\STATE Evaluate the local objective function $F(S_i,a_i)$, generate reward $R(S_i,a_i)$
		\STATE Update the $s$ DNNs weights $\theta$
		\ENDIF
		\ENDFOR
		\STATE 	Output DNN parameter $\theta$ as meta-parameter $\psi$
	\end{algorithmic}  
\end{algorithm}

\section{Performance Evaluation}
\label{sec:evaluation}
In this section, we evaluate the performance of the proposed DMRO framework for solving the optimization problem of offloading decision-making under different MEC environments. 

\subsection{Simulation setup}
In our simulation, we assume that there are four IoT users, and each user has five workflows. The size of the first subtask of each workflow is $50-100$ MB, and the size of subsequent tasks is $10-50$ MB. The amount of computation for each subtask is $10^3-10^5$ MHz randomly distributed. For the DNN structure, we consider a fully connected DNN consisting of one input layer, two hidden layers, and one output layer in the proposed DMRO framework. The parameters $\alpha$ and $\beta$ are both set to $1$.
In addition, we set the environmental information as listed in Table~\ref{tab:parameter}.
\begin{table}[ht]
\centering
\caption{Environmental Information}
	\label{tab:parameter}
	\renewcommand{\arraystretch}{1.25}
\begin{tabular}{|c| c|c|c|c|c|}
	\hline
	$C_{local}$ & 30 \text{MHz} &$B_{0,1}$ & $800$ \text{MB/s}&$d_{local}$ & $0.3$ \text{J/MB}\\
	\hline
	$C_{edge}$ & $70$ \text{MHz}&$B_{1,2}$ & $200$ \text{MB/s}&$d_{edge}$ & $0.15$ \text{J/MB}\\
	\hline
	$C_{cloud}$ & $150$ \text{MHz}&$B_{0,2}$ & $10$ \text{MB/s}&$d_{cloud}$ & $0.1$ \text{J/MB}\\
	\hline
\end{tabular}
\end{table}

\subsection{Convergence Performance}

 \begin{figure}[htbp!]
	\centering
	\includegraphics[width=3.6in,trim=0.3in 0in 0in 0in]{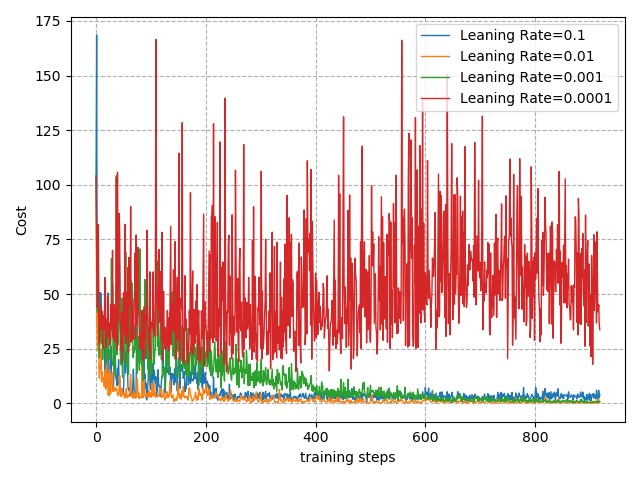}
\caption{Convergence performance under different learning rates}
\label{Convergence_learningrate}
\end{figure}  

\begin{figure*}[htbp]
\centering
	\subfigure[Batch size 128]{
			\includegraphics[width=1.7in,trim=0.1in 0in 0.1in 0.1in]{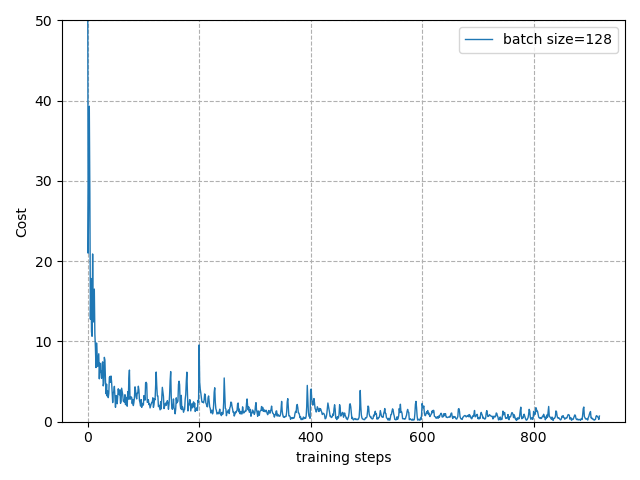}}
	\subfigure[Batch size 256]{   	
			\includegraphics[width=1.7in,trim=0.1in 0in 0.1in 0.1in]{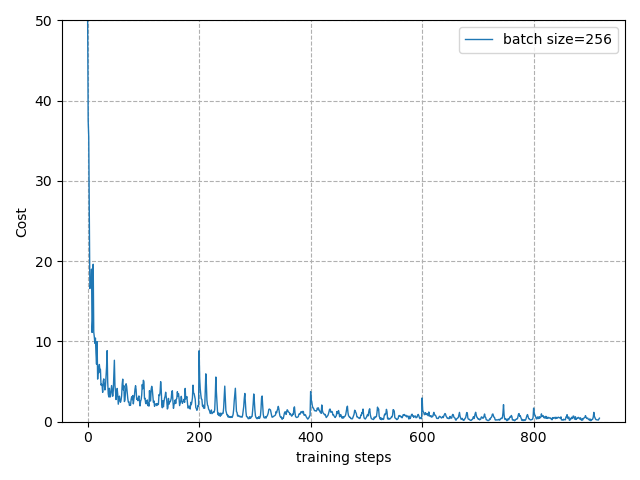}}
	\subfigure[Batch size 512]{ 
			\includegraphics[width=1.7in,trim=0.1in 0in 0.1in 0.1in]{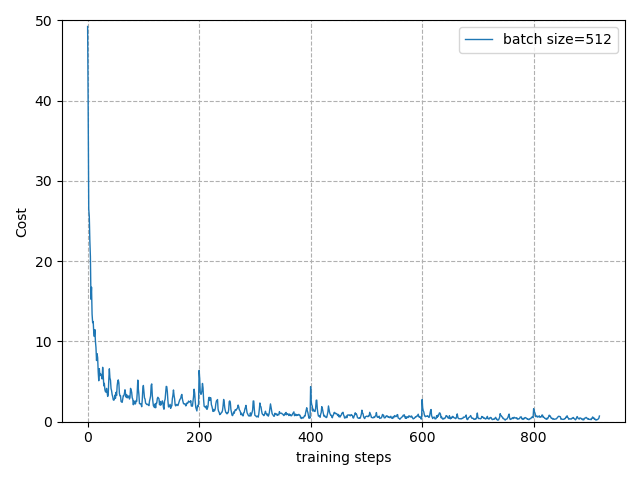}
			} 
	\subfigure[Batch size 1024]{
			\includegraphics[width=1.7in,trim=0.1in 0in 0.1in 0.1in]{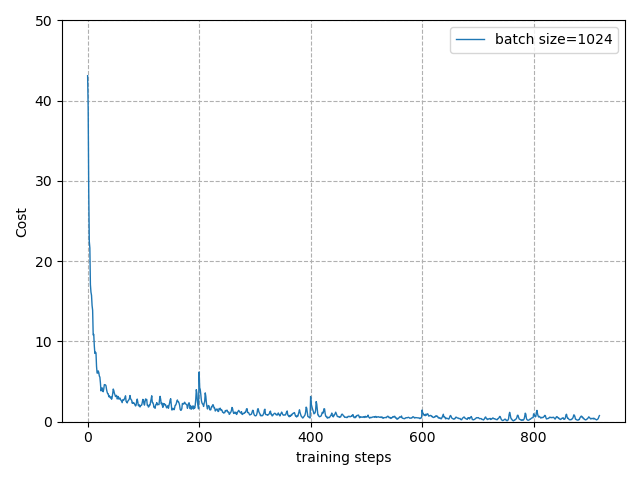}
	}
\caption{Convergence performance under different batch sizes}
\label{Convergence_batchsize}
\end{figure*}

Figure~\ref{Convergence_learningrate} shows the convergence performance of our model under different learning rates, where the abscissa is the number of training steps and the ordinate is the loss of the neural network. It can be found that when the learning rate is too low, it will not be able to converge. However, when the learning rate is 0.01, the convergence effect is the best, so we will use a learning rate of 0.01 in the next experiment. 

Figure~\ref{Convergence_batchsize} shows the convergence performance of our model under different batch sizes. It can be seen that the batch size has less effect on the convergence, but as the batch size increases, the volatility of the curve becomes smaller. It is worth noting that there is a small fluctuation in the curve every 200 steps, which is mainly due to the parameter freezing mechanism. In the model, we set the network parameters to the target network every 200 steps. As a result, the parameters will fluctuate every 200 steps, but it does not affect the convergence of the model.

%\subsection{System Utility}

\subsection{Comparison Experiments}

To gain insight into the proposed DMRO scheme for edge offloading decision, the following methods are implemented for comparison: 
\begin{itemize}
	\item \emph{Local-only scheme}: In this method, all tasks of workflows are executed locally on the IoT device. The results of this method can be used as a benchmark to analyze the gain of task offloading techniques.
	 
	\item \emph{Edge-only scheme}: This is a full offloading scheme. In this method, all tasks of workflows are fully offloaded to the edge servers for further processing.      
	
	\item \emph{Cloud-only scheme}: This is a full offloading scheme. In this method, all tasks of workflows are fully offloaded to the cloud server for further processing.      
	
	\item \emph{Deep Q-Network scheme}: This is a partial offloading scheme based on the Deep Q-Network algorithm \cite{haghighi2018offloading}, where it can be regarded as a simplified DMRO algorithm with only one parallel network. In this method, we use the Deep Q-Network in making dynamic offloading decisions. 
	
	\item \emph{DMRO scheme}: This is a partial offloading scheme based on the proposed DMRO scheme. It is designed to find the optimal offloading scheme that minimizes the weighted delay and energy consumption. 
\end{itemize} 
\begin{figure}[htbp]
	\centering
	\includegraphics[width=3.3in, trim=0.2in 0.15in 0.2in 0.15in]{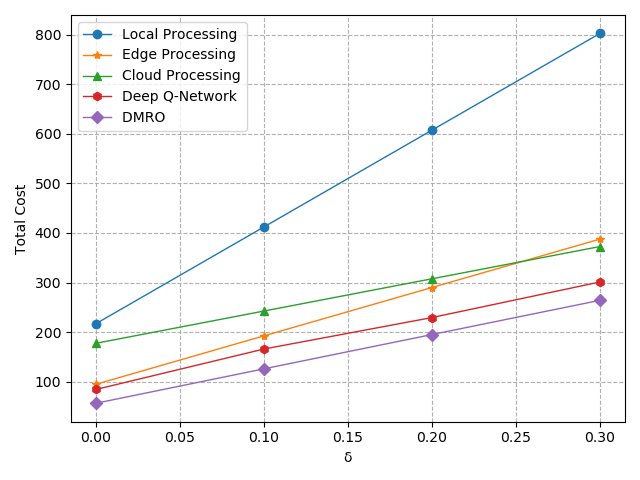}
	\caption{Comparison of different offloading schemes under different weights}
	\label{7}   
\end{figure}  

The comparison results are as shown in Fig.~\ref{7}, where the abscissa is the weight ratio of delay to energy consumption, and the ordinate is the value of the objective function. Especially, when the weight value is 0, it means that only delay is considered. The figure shows that the DMRO algorithm can achieve the minimum total cost among the five methods, and the DQN algorithm and the DMRO algorithm have the same trend, which are better than the \emph{local-only scheme}, the \emph{edge-only scheme} and the \emph{cloud-only scheme}. In addition, as the weight ratio of energy consumption increases, the total consumption of local execution increases rapidly, which also meets our expectations, indicating that local devices are more sensitive to energy consumption.

\subsection{Fast Learning}
We show the effect of the proposed meta algorithm in fast offloading decision learning under different IoT environments. We first set up two types of environments, i.e., the training environment and the testing environment, as shown in Table~\ref{tab:parameterValues}.
\begin{table}[ht!]
	\caption{Evaluation Parameters}
	\label{tab:parameterValues}
	\renewcommand{\arraystretch}{1.25}
	\centering
	\begin{tabular}{|c| c|c|c|}
		\hline
		\multicolumn{2}{|c|}{train}&\multicolumn{2}{|c|}{test}\\
		\hline		
		$C_{local}$ & $15-25$ \text{MHz}& $C_{local}$ & $30$ \text{MHz}\\
		\hline
		$C_{edge}$ & $50-60$ \text{MHz} & $C_{edge}$ & $70$ \text{MHz}\\
		\hline
		$C_{cloud}$ & $160-170$ \text{MHz} & $C_{cloud}$ & $150$ \text{MHz}\\
		\hline
		$B_{0,1}$ & $600-700$ \text{MB/s} & $B_{0,1}$ & $800$ \text{MB/s}\\
		\hline
		$B_{1,2}$ & $100-150$ \text{MB/s} & $B_{1,2}$ & $200$ \text{MB/s}\\
		\hline
		$B_{0,2}$ & $20-30$ \text{MB/s} & $B_{0,2}$ & $10$ \text{MB/s}\\
		\hline
		$d_{local}$ & $0.3$ \text{J/MB} & $d_{local}$ & $0.3$ \text{J/MB}\\
		\hline
		$d_{edge}$ & $0.15$ \text{J/MB}& $d_{edge}$ & $0.15$ \text{J/MB}\\
		\hline
		$d_{cloud}$ & $0.1$ \text{J/MB} & $d_{cloud}$ & $0.1$ \text{J/MB}\\
		\hline
		
	\end{tabular}
\end{table}

\begin{figure}[ht!]
	\centering
	\includegraphics[width=3.3in, trim=0.2in 0.15in 0.2in 0.15in]{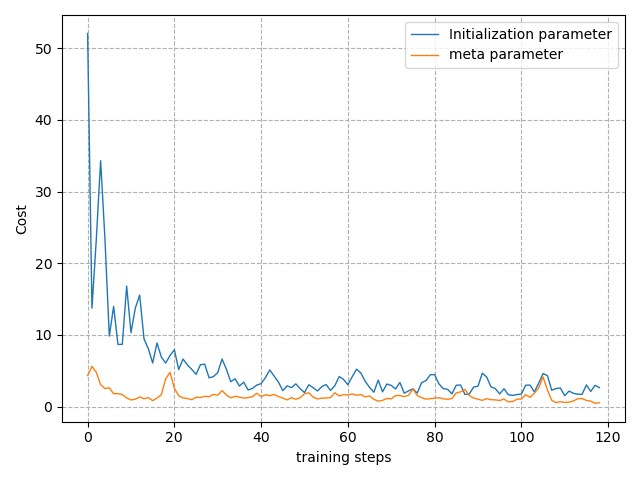}
	\caption{Convergence performance by meta algorithm.}
	\label{8}   
\end{figure}

We input the trained meta parameters into the test environment. Figure~\ref{8} shows the comparison between the meta parameters and the initialized parameters in terms of convergence. It can be seen that the initial convergence of meta parameters is better than the traditional initialization parameters. Besides, it illustrates that networks using meta parameters can converge faster.

\begin{figure}[ht!]
	\centering
	\includegraphics[width=3.3in, trim=0.2in 0.15in 0.2in 0.15in]{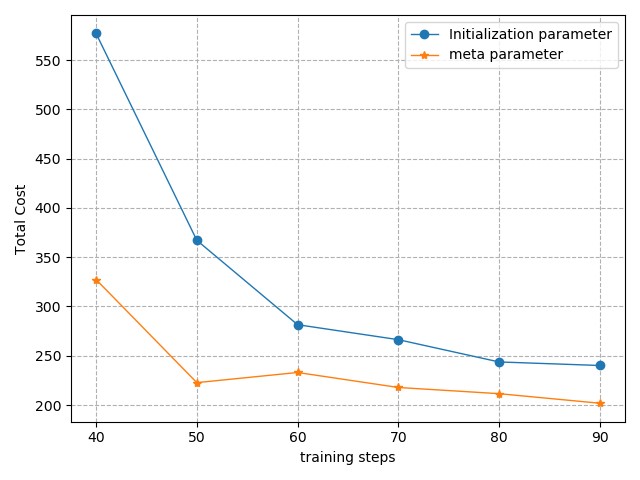}
	\caption{The performance of meta parameters and initialized parameters}
	\label{9}   
\end{figure}

Figure~\ref{9} shows the comparison of the effect of meta parameters and initialized parameters on the total cost. It can be seen from the figure that the decision result of the neural network using meta parameters is significantly better than that of the traditional initialization parameter network, and a low-cost offloading decision can be given after a few rounds of training. Therefore, by learning the initial parameters of the neural network, the offloading decision model can quickly adapt to the new environment. In addition, it is worth noting that although the convergence degree of the network does not change much during the process of training 40-80 steps, the task offloading decision model has been optimized, and then a more reasonable offloading solution is given.

\section{Conclusion}
\label{sec:conclusion}
This paper has proposed an edge offloading framework to deal with the task offloading decision-making in heterogeneous IoT-edge-cloud computing environments. The DMRO framework includes a task offloading decision model based on distributed deep reinforcement learning algorithm and a training initial parameter model based on deep meta-learning, which aims to solve the problem of poor portability of neural networks. 

Experimental results show that the DMRO framework has a better effect on task offloading decisions than full offloading methods and conventional reinforcement learning-based methods. In addition, due to the use of meta parameters, the model has stronger portability and  rapid environment learning ability. Once the MEC environment changes, the model can quickly converge, and only a small number of learning steps are needed to give low-cost offloading solutions. We expect that the initial parameters can be changed adaptively in response to environmental parameters in future work.

Although this paper only focuses on the scenario with one edge server and one cloud server, this model is highly scalable and can be easily extended to other scenarios with multiple edge servers and cloud servers. In the future, we plan to implement the proposed DMRO framework for intelligent offloading when considering hardware platforms, e.g., graph-based Network Interface Controller (NIC) offload and device offload \cite{Diaz2019,Shi2019,Hui2020}.

\ifCLASSOPTIONcompsoc
	\section*{Acknowledgments}
	\else
	\section*{Acknowledgment}
	\fi
	
This work is partially supported by the National Natural Science Foundation of China (61801325) and the Natural Science Foundation of Tianjin City (18JCQNJC00600). 	
 
\ifCLASSOPTIONcaptionsoff
  \newpage
\fi

\bibliographystyle{IEEEtran}
\bibliography{references}

\vspace{1cm}

\begin{IEEEbiography}[{\includegraphics[width=1.05in,trim=0in 0.7in 0in 0.4in,clip,keepaspectratio]{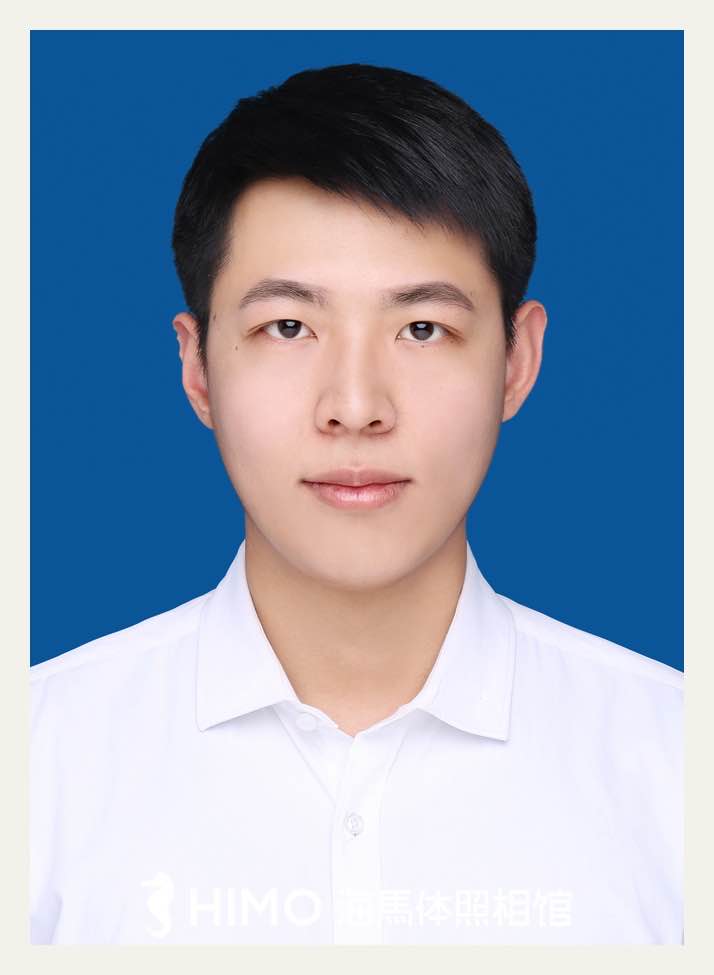}}]{Guanjin Qu}
	 is currently working toward the Master degree at the Center for Applied Mathematics, Tianjin University, China. His research interests include distributed deep learning and edge computing.
	\end{IEEEbiography}

\begin{IEEEbiography}[{\includegraphics[width=1.05in,trim=0in 0.7in 0in 0.3in,clip,keepaspectratio]{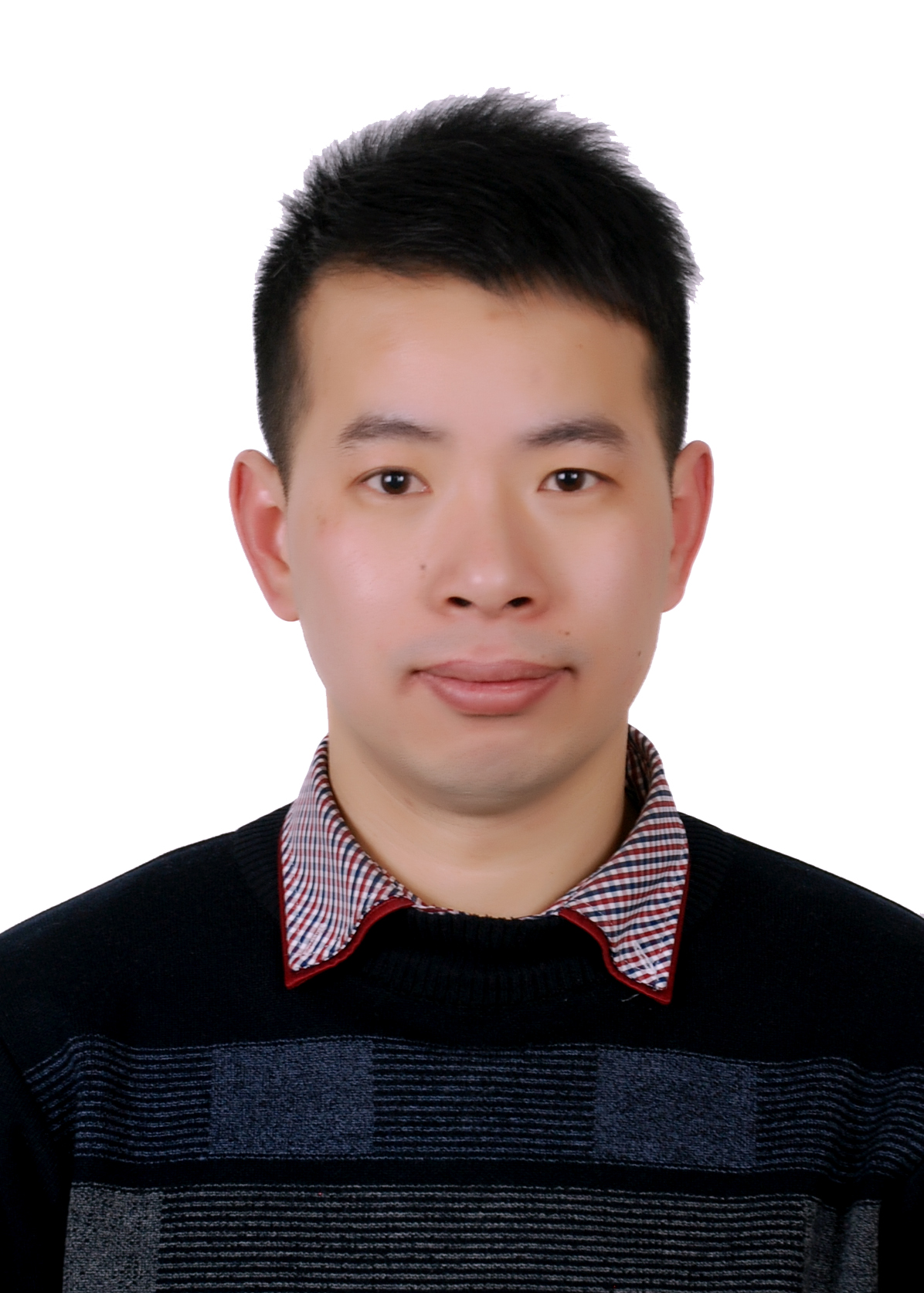}}]{Huaming Wu}
received the B.E. and M.S. degrees from Harbin Institute of Technology, China in 2009 and 2011, respectively, both in electrical engineering. He received the Ph.D. degree with the highest honor in computer science at Freie Universit\"at Berlin, Germany in 2015. He is currently an associate professor at the Center for Applied Mathematics, Tianjin University, China. His research interests include model-based evaluation, wireless and mobile network systems, mobile cloud computing and deep learning.
\end{IEEEbiography} 
%	\vspace{-1.5cm}

\end{document}